# Improvement of EAST Data Acquisition Configuration Management

Chen Ying, Li Shi

*Abstract*—The data acquisition console is an important component of the EAST data acquisition system which provides unified data acquisition and long-term data storage for diagnostics. The data acquisition console is used to manage the data acquisition configuration information and control the data acquisition workflow. The data acquisition console has been developed many years, and with increasing of data acquisition nodes and emergence of new control nodes, the function of configuration management has become inadequate. It is going to update the configuration management function of data acquisition console. The upgraded data acquisition console based on LabVIEW should be oriented to the data acquisition administrator, with the functions of managing data acquisition nodes, managing control nodes, setting and publishing configuration parameters, batch management, database backup, monitoring the status of data acquisition nodes, controlling the data acquisition workflow, and shot simulation data acquisition test. The upgraded data acquisition console has been designed and under testing recently.

## I. INTRODUCTION

EAST (Experimental Advanced Superconducting Tokamak) has been in operation since 2006 and the physics experiments have been developed further in support of high-performance steady-state operation [1]. Similar to other fusion facilities [2-5], EAST established the data acquisition system to provide unified diagnostic data acquisition. The data acquisition console is used to manage the configuration information and control the data acquisition workflow as an integral component of the EAST data acquisition system.

With the development of EAST experiments, the Data Acquisition nodes (DAQ nodes) are increasing and some control nodes are emerging. The current data acquisition console was developed many years ago and designed only for unified diagnostic data acquisition. It is planned to update the data acquisition console, mainly to upgrade the configuration management function. Key requirements of the upgraded Data Acquisition console (DAQ console) are described as follows:

1) There are about 60 DAQ nodes in the EAST data acquisition system, which can totally provide more than 3,000 channels for different diagnostics in different physic locations. The upgraded data acquisition console should still manage these DAQ nodes.

2) The current DAQ console was only designed for diagnostic data acquisition, without considering the later control nodes. Therefore, the upgraded data acquisition console needs to facilitate the management of these control nodes.

3) Now it is not convenient to add or delete a node, and users need to manually modify the relevant records one by one. Additionally, there is no function of configuration database backup. Therefore, it is necessary to optimize the configuration management function and provide the database backup function.

4) In order to facilitate the internal data acquisition test, it is necessary to add the function of the shot simulation data acquisition test.

5) EAST aims at long-pulse and high-performance operation. The upgraded DAQ console should retain the function of automatically and continuously controlling the long-pulse data acquisition workflow.

## II. SYSTEM DESIGN

Fig.1 illustrates the architecture of the current EAST data acquisition system, which consists of a data acquisition console, the data server cluster, DAQ nodes, and control nodes. All the components are in a Local Area Network (LAN).

DAQ nodes and control nodes are distributed in different physical positions. The main functions of the DAQ console are managing all nodes' configuration information and controlling data acquisition workflow.

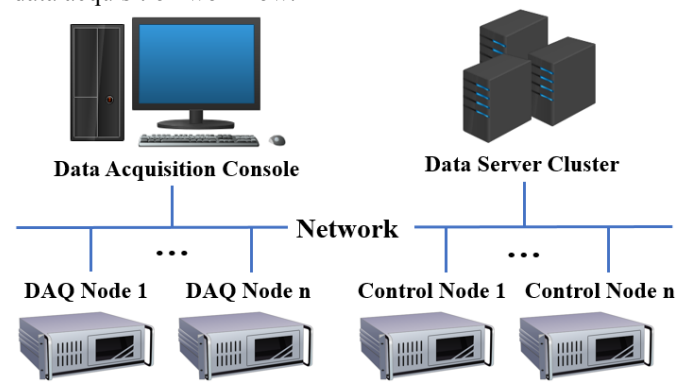

Fig. 1. Data Acquisition System Architecture

As shown in Fig. 2, according to different functions, the DAQ console can be divided into two modules: the management module and the DAQ control module. The

Manuscript received October 29, 2020. This work was supported by the National MCF Energy R&D Program of China under Grant 2018YFE0302100.

Chen Ying and Li Shi are with the Institute of Plasma Physics, Chinese Academy of Sciences, Hefei, 230031, China (e-mail: cheny@ipp.ac.cn; lishi@ipp.ac.cn).

improvement is mainly in the management module, while the data acquisition control logic of the DAQ control module is not changed.

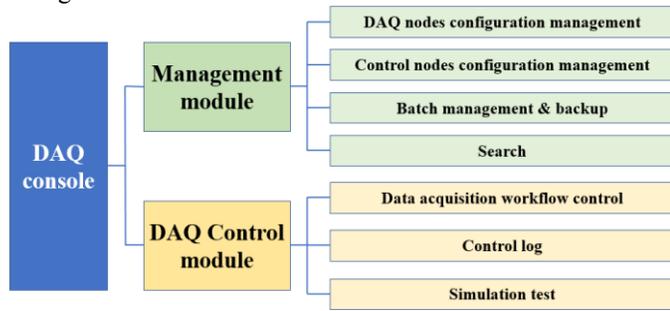

Fig. 2. DAQ console modules

In the management module, it improves the configuration management of DAQ nodes, adds the management of control nodes, adds the function of batch management and configuration database backup with excel file, and retains the function of searching signal information. For DAQ nodes, the configuration information includes status information, data storage information, card information, and channel information. The status information describes the DAQ device information. The data storage information describes the corresponding storage server information. The card information describes the configuration information of each data acquisition card. The channel information describes the corresponding signal information of each channel. For control nodes, the configuration information includes the node name, IP, etc.

In the control module, it can automatically control the active DAQ nodes' workflow, which is consistent with the original system. Compared with the original system, it adds the functions of recording and querying the control log information, and shot simulation test.

III. SYSTEM IMPLEMENTATION

A. Management module

The management module provides a Graphical User Interface (GUI) as shown in Fig.3. The GUI is developed with LabVIEW and has three parts: operation area, display area, and control area. In the operation area, the user can perform some operations to view and manage the nodes' information. Fig.4 shows the operation details with five classifications: contents, index, search, log, and operation.

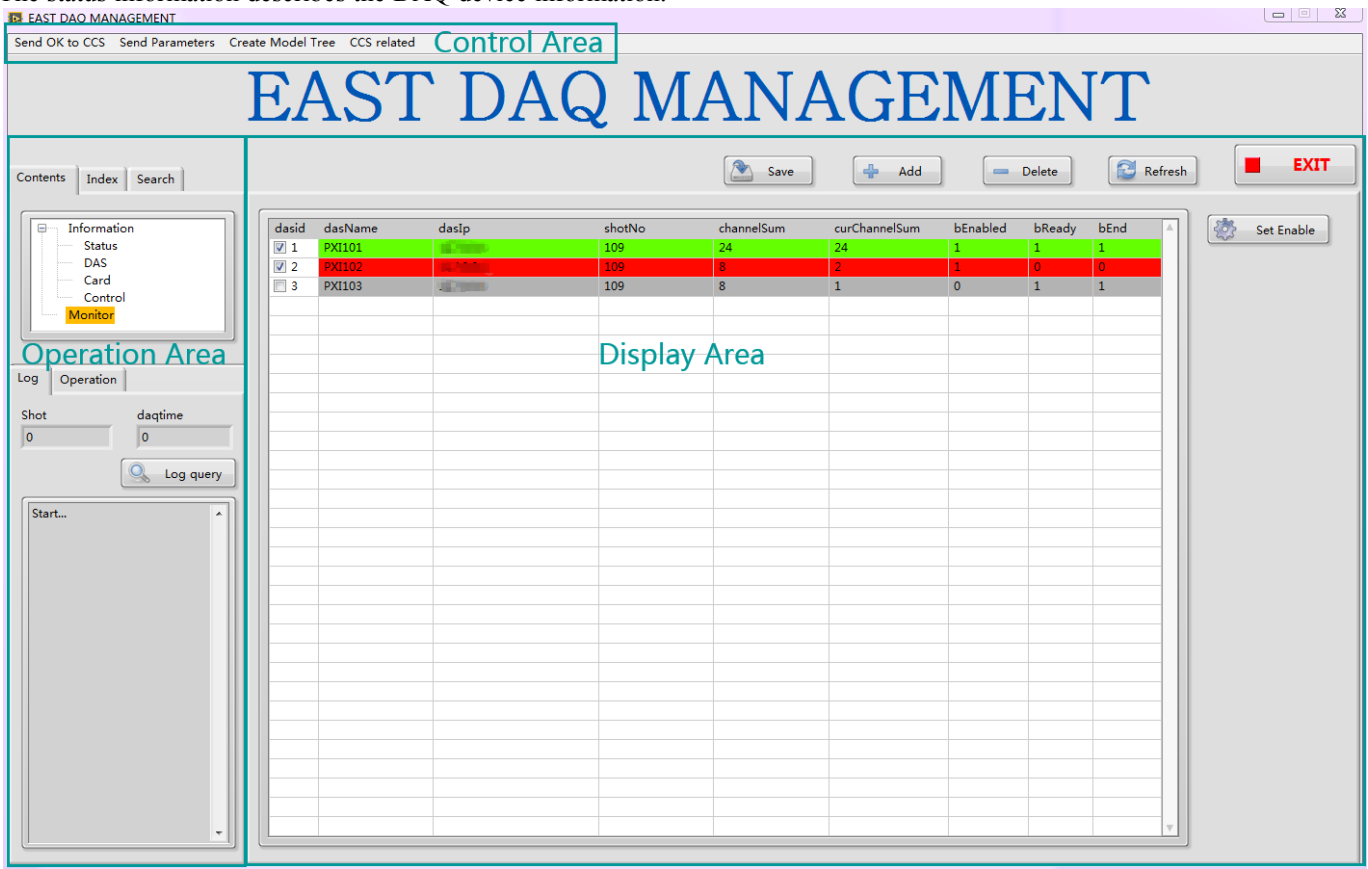

Fig. 3. The graphical user interface of the management module

In the contents classification, there is a tree list showing five kinds of information. The Status, DAS, and Card information are used to describe DAQ nodes, including the overall information of each data acquisition device, the data storage information of each DAQ node, and the configuration details of each data acquisition card. The Control information is used to describe control nodes. The Monitor information can show the status of all DAQ nodes clearly in different colors. All information will be shown in the display area in form of tables, and the user can edit these data.

In the index classification, the user can get the related information of the DAQ node such as details of card or channel information. The user can also get some channel information in the search classification with link name (the name of the signal cable) or signal name.

In the log classification, it shows the current shot number，data acquisition duration, and control log information. After clicking the log query button, the user can search the log information in the display area with a specific date or shot number range.

update the new configuration with the second menu bar. EAST uses MDSPlus to store the experimental data, so the module provides the third menu bar for manual creating model tree. Finally, the last menu bar is set for data acquisition internal tests.

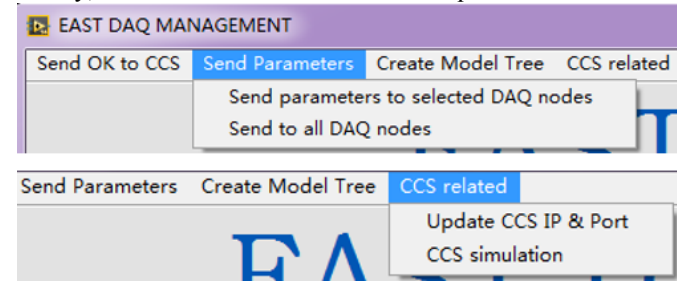

Fig. 5. Control area

## IV. CONCLUSION

The upgraded data acquisition console is an essential component of EAST data acquisition system. The upgraded data acquisition console has been designed and implemented a beta version. The beta DAQ console is based on LabVIEW and retains the function of the original console, which can manage the DAQ nodes and control the long-pulse data acquisition workflow, with adding some new functions on this basis. It adds functions of managing control nodes, batch managing the configuration information, configuration database backup, and recording log. At present, the beta DAQ console is considered to be used for EAST data acquisition system internal test, especially for the pre-launch testing of a new DAQ node.

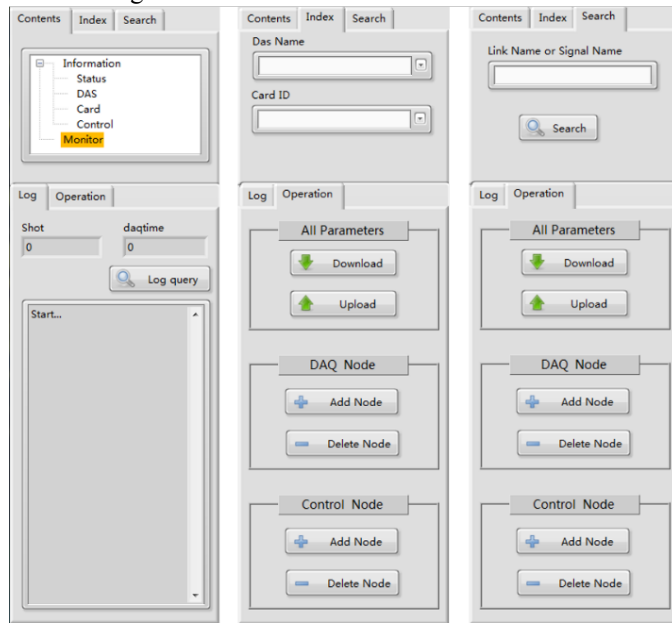

Fig. 4. Operation area

In the operation classification, if the user wants to perform batch operations of configuration information easily, the user can export all parameters into a excel file from the configuration database with the download button, and performs batch operations in the excel file, such as add, delete or update operation, then import configuration information from the excel file into configuration database with the upload button. Meanwhile, the user can add DAQ nodes or control nodes with add node buttons and directly delete all information of selected DAQ nodes or control nodes via delete node buttons without deleting all relevant records one by one.

### B. DAQ control module

The DAQ control module can automatically control the enabling DAQ nodes' workflow, which is consistent with the original system. As described in section A and Fig.3, the user can choose the Monitor information in the contents classification to view the current status of each DAQ node in the display area. It uses gray for inactivity, green for waiting, and red for data acquiring or abnormal. As shown in Fig.5, it provides some functions for the user in the control area. In case of the data acquisition fault, the user can use the first menu bar after troubleshooting to notify the CCS that the data acquisition system returns to normal. When some configuration information has been changed, the user can notify nodes to